# A Nuclear Physics Program at the ATLAS Experiment at the CERN Large Hadron Collider

S. Aronson, K. Assamagan, H. Gordon, M. Leite, M. Levine, P. Nevski, H. Takai, and S. White
**Brookhaven National Laboratory**

B. Cole
**Columbia University**

J. Nagle
**University of Colorado at Boulder**

## Abstract

The ATLAS collaboration has significant interest in the physics of ultra-relativistic heavy ion collisions. We submitted a Letter of Intent to the United Stated Department of Energy in March 2002. The following document is a slightly modified version of that LOI. More detailed rate calculations and physics capabilities, in addition to the presentations associated with the LOI, can be found at the ATLAS Heavy Ion web page:

http://atlas.web.cern.ch/Atlas/GROUPS/PHYSICS/SM/ions

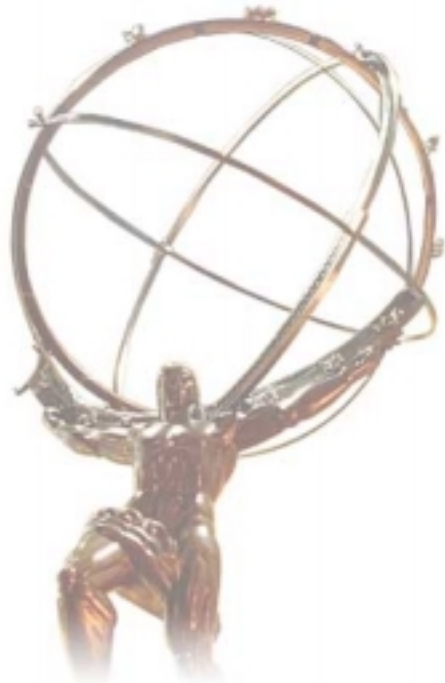

# 1. Introduction

The ATLAS detector has been designed for determining the source of electroweak symmetry breaking using proton-proton collisions. Many of its components could serve to measure the properties of nucleus-nucleus and proton-nucleus collisions: the silicon pixel and silicon strip tracking elements close to the interaction point, finely subdivided electromagnetic calorimetry and a muon spectrometer that does not depend on tracking inside the calorimeters. We intend to study the physics potential of the ATLAS detector in nucleus-nucleus and proton-nucleus collisions as outlined in this note.

Lead-Lead collisions at LHC will take place at over a 1,000 TeV total center-of-mass energy, which is more than an order of magnitude higher in center of mass energy than achievable at the Relativistic Heavy Ion Collider (RHIC). It is expected that energy densities of up to 30 GeV/fm$^3$ can be reached allowing for the exploration of forms of matter not previously seen in the laboratory.

We believe that the ATLAS detector has enormous capabilities for both nucleus-nucleus and proton-nucleus physics. The ATLAS nuclear physics program will have some aspects that are competitive with what ALICE or CMS can achieve. Some overlap with measurements from ALICE and CMS is expected. The diversity of goals among all experiments is crucial for the overall success of the LHC nuclear physics program.

The construction of the ATLAS detector is well under way, and for a relatively small cost to the nuclear physics community, state of the art measurements could be made to elucidate the nature of nuclear matter at the highest possible energies and parton densities. Here we outline several interesting physics goals. Detailed simulation will need to be done to assure that these goals can be met. This letter of intent is just the beginning of the process.

# 2. The ATLAS detector

The ATLAS detector is designed to study proton-proton collisions at the LHC design energy of 14 TeV in the center of mass. The physics pursued by the collaboration is vast and includes: Higgs boson search, searches for SUSY, and other scenarios beyond the Standard Model. To achieve these goals at a full machine luminosity of $10^{34}$ cm$^{-2}$s$^{-1}$, the calorimeter is designed to be as hermetic as possible and has extremely fine grain segmentation. The detector as shown in Figure 1 is a combination of three subsystems: Inner tracker system, electromagnetic and hadronic calorimeters and full coverage muon detection.

The inner detector system is composed of (1) a finely segmented silicon pixel detector, (2) Semiconductor Tracker (SCT) and (3) the Transition Radiation Tracker (TRT). The segmentation is optimized for proton-proton collisions at design machine luminosity.

The calorimeter system in the ATLAS detector is divided into electromagnetic and hadronic sections and covers pseudo-rapidity $|\eta|<4.9$. The EM calorimeter is an accordion liquid argon device and is finely segmented longitudinally and transversely for $|\eta|<3.1$. The first longitudinal segmentation has a granularity of $\Delta\eta \times \Delta\phi$ = 0.003 x 0.1 in the barrel and slightly coarser in the endcaps. The second longitudinal segmentation is composed of $\Delta\eta \times \Delta\phi$ = 0.025 x 0.025 cells and the last segment $\Delta\eta \times \Delta\phi$ = 0.05 x 0.05 cells. In addition a finely segmented $\Delta\eta \times \Delta\phi$ =0.025x0.1 pre-sampler system is present in front of the EM calorimeter. The overall energy resolution of the EM calorimeter determined experimentally is $10\%/E^{1/2} \oplus 0.5\%$. The calorimeter also has good pointing resolution (60 mrad/$E^{1/2}$) for photons and timing resolution better than 200 picoseconds for showers of energy larger than 20 GeV.

The hadronic calorimeter is also segmented longitudinally and transversely. Except for the endcaps, the technology utilized for the calorimeter is a lead-scintillator tile structure with a granularity of $\Delta\eta \times \Delta\phi$ = 0.1 x 0.1. In the endcaps the hadronic calorimeter is implemented in liquid argon technology for radiation hardness with the same granularity as the barrel hadronic calorimeter. The energy resolution for the hadronic calorimeters is $50\%/E^{1/2} \oplus 2\%$ for pions. The very forward region, up to $|\eta|<4.9$ is covered by the Forward Calorimeter implemented as an axial drift liquid argon calorimeter. The overall performance of the calorimeter system is described in [1].

The muon spectrometer in ATLAS is located behind the calorimeters, thus shielded from hadronic showers. The spectrometer is implemented using several technologies for tracking devices and a toroidal magnet system. Most of the volume is covered by MDTs, (Monitored Drift Tubes). The forward region where the rate is high, Cathode Strip Chamber technology is chosen. The stand-alone muon momentum resolution is of the order of 2% for muons with $p_T$ in the range 10 - 100 GeV.

The performance of each subsystem is summarized in a series of Technical Design Reports [2].

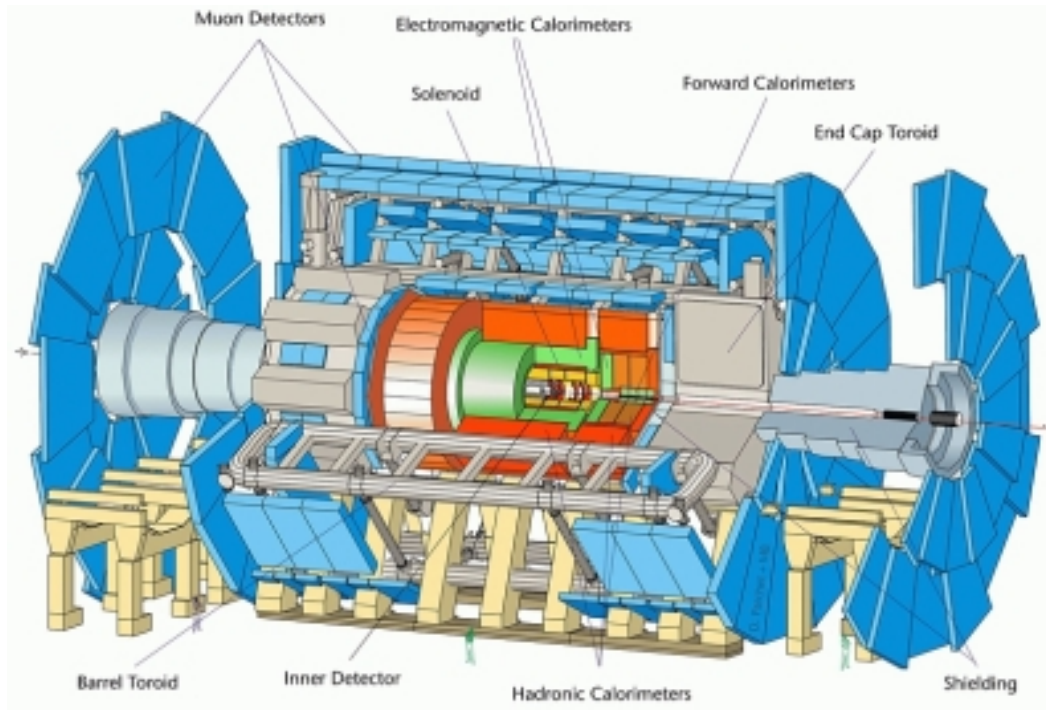

**Figure 1 The ATLAS detector.**

## 3. The ATLAS nuclear physics program

The ATLAS experiment has the optimal electromagnetic and hadronic calorimetry among all experiments at the LHC for performing energy measurements. In central Pb-Pb reactions, while the inner silicon layers will provide good track information, the transition radiation tracker (TRT) will have very low efficiency due to high occupancy. The calorimeter has the best granularity compared to other LHC detectors with excellent energy and timing resolution and will be fully functional. This will provide the optimal opportunity for measuring jets using detailed shape information. The muon spectrometer can be used for b-jet tagging as well. The proposed nuclear physics program with the ATLAS detector is outlined as follows:

- Global variable measurements – Measurements of total transverse energy flow $E_T$ and $dE_T/d\eta$, event multiplicity N and $dN/d\eta$, and elliptic flow will provide an important measurement of the energy density, reaction dynamics, and will permit the selection of the collision impact parameter. These measurements can be done extremely well with the ATLAS detector.
- Jet quenching studies – Hot QCD matter may modify jet properties (e.g. cone radius and energy) because partons may radiate soft gluons in the presence of a dense quark gluon plasma before hadronization. Early RHIC results suggest that quenching may be the cause for the suppression of hadrons at large $p_T$. At LHC jet properties can be better measured when compared to RHIC due to the higher collision energy.

We are exploring the feasibility of identifying γ-jet, jet-jet, and perhaps Z-jet channels in the presence of the heavy ion soft background.
- <u>Heavy quarks</u> – Theoretical work indicates that heavy quarks propagating through the QCD medium lose much less energy via gluon bremsstrahlung than light quarks, and results in less quenching. b-jets in ATLAS, for example, could be tagged by the associated muon and measure this quenching.
- <u>Quarkonia</u> – Screening of the long- range attractive potential is expected within the hot QCD medium. The observation of ϒ states suppression is a direct consequence of hot plasma formation. We are in the process of evaluating the ϒ mass resolution attainable with the ATLAS muon spectrometer.
- <u>Proton-nucleus collisions</u> – Proton-nucleus physics at the LHC will focus on the perturbative production of gluons and the modifications of the gluon distribution in the nucleus at low x. We believe that there is an excellent opportunity to study proton-nucleus physics with ATLAS. It will also provide a baseline for the understanding of nucleus-nucleus physics.
- <u>Nucleus-nucleus ultra peripheral collisions</u> – The highly Lorentz contracted electric field of heavy nuclei is a source of high energy photon-photon and photon nucleon collisions. Heavy quark photo-production may allow for the determination of the saturation scale within the context of color glass condensate model.

## 4. ATLAS Physics in Nucleus-Nucleus Collisions

The nucleus-nucleus physics program in ATLAS will make extensive use of the large calorimetric coverage, the muon spectrometer, the pixel detector, and the SCT detector system. The coverage of the ATLAS detector will permit the event characterization and direct measurement of the energy density for nucleus-nucleus collisions. In the early stages of the heavy ion runs, measurements of the total energy flow, $E_T$, and its dependence on pseudorapidity $dE_T/d\eta$ will be possible. The silicon pixel system could also be used for the measurement of total charged particle multiplicity and also $dN/d\eta$. An analysis similar to what was done by the PHOBOS collaboration could be done. The calorimeter will also provide a measure of the reaction plane in non-central collisions via an azimuthal angle energy distribution.

In ultra-relativistic central collisions (b ~ 0 fm), an enormous number of virtual partons, dominated by gluons, are freed from the nuclear wave function. These partons should form a quark-gluon-plasma characterized by quark de-confinement and restoration of approximate chiral symmetry. One major advantage of the higher LHC collision energy compared to RHIC is that the initial parton densities are so high that a description of the gluons as color field solutions to the Yang-Mills equation may be relevant. This picture, referred to as the color glass condensate, has had some initial success in describing RHIC data, but would be in a more clearly relevant regime of $Q^2$ at the LHC [3,4].

At the LHC energy regime, hard processes namely jets can serve as probes of the hot QCD matter created after the nuclei collide. Because of the expected longer lifetime and higher temperature of the de-confined state these probes could be a direct evidence of the quark interaction with the plasma. One of the key observables is the measurement of parton probes of the plasma medium via their induced gluon radiation, often referred to as jet quenching [5-7]. The parton energy loss is directly related to the initial gluon density of the system, which is expected to be over a factor of ten higher than at RHIC. As these partons travel through the plasma their energy loss should result in a softening of the final jet fragmentation into hadrons. Given the large calorimetric coverage and the fine segmentation of the ATLAS detector, these probes are the most suitable to be used in the nuclear physics program.

Early results from the PHENIX experiment at RHIC [8] have shown first hints of energy loss effects. They also point out the critical need to measure fragmentation functions with identified particles and with known original total jet energy. In Figure 2 the suppression factor $R_{AA}$ for identified neutral pions is shown as function of the transverse momentum from the PHENIX experiment. Overlayed is a theoretical calculation from Vitev et al.[9] that shows qualitative agreement at RHIC energies and show predictions for LHC energies. It is noteworthy that there is larger suppression due to the expected higher gluon density and that the suppression remains significant for $p_T > 50$ GeV/c. Detailed simulation studies will be required to understand ATLAS capabilities of detecting $\pi^0$ and $\eta$ up to large values of $p_T$ (~50 GeV).

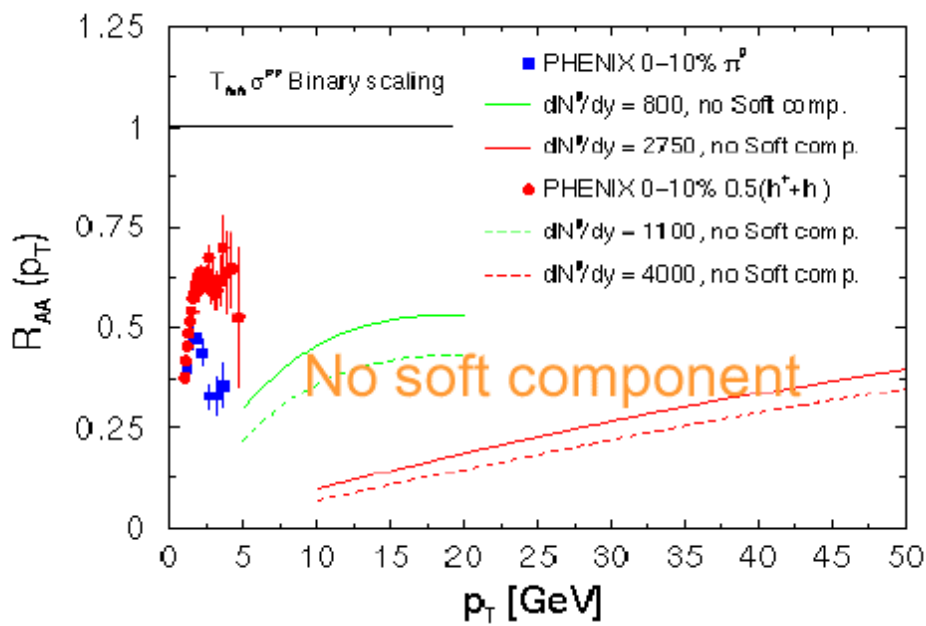

**Figure 2 PHENIX experimental data for $R_{AA}$ = particle invariant yield in A-A collisions divided by the same particle yield in proton-proton collisions scaled up by the expected number of binary collisions. The theoretical curves for RHIC and LHC from Vitev et al. [9] are also shown.**

There is excellent opportunity in ATLAS to measure γ-jet, jet-jet and Z-jet events where one can more fully characterize the modified fragmentation functions. In particular, the γ(or Z) in γ(or Z)-jet processes provides a "control" over the away-side jet energy and direction that will allow the physics of quenching to be studied quantitatively and in great detail [10]. The effects of hard gluon radiation on the photon/jet energy imbalance and angular distribution can be studied in great detail using the high-statistics p-p data set. The γ-jet channel requires the identification of a photon. In proton-proton collisions the rejection of $\gamma/\pi^0$ is about a factor of three up to a $p_T$ of 50 GeV. However, the heavy ion environment presents considerable more challenge. $Z^0$ production rates have been estimated by Wang and Huang [10]. For $p_T$ larger than 40 GeV, we expect of the order of 500 $Z^0 \rightarrow \mu\text{-}\mu\text{+}$ events for one month-run.  Therefore multiple runs may be required to extract relevant information on jet fragmentation.

Recent theoretical investigations [11] have indicated that charm and bottom quarks propagating through a dense partonic medium will have a supressed gluon radiation.  Thus the possibility of measuring b-jets in ATLAS would give an important comparison measurement to the light quark and gluon jets. Such measurements would have important implications on gluon shadowing and saturation models. Tagging of b-jets by the associated muon is possible in the proton-proton environment [1]. We are currently studying the possibility of tagging b-jets by matching a measured muon in the standalone muon spectrometer to the jet measured by the calorimeter system in the heavy ion environment.  In the same spirit we are also investigating the possibility of probing the media with the heaviest of the quarks: the top quark.

It is worthwhile commenting that reconstruction of jet with energies above 40 GeV seems feasible in the ATLAS environment. Below this energy the large combinatorial background will greatly compromise jet energy measurements. However filter algorithms based on jet profile libraries measured in proton-proton collisions may substantially improve measurements.

Suppression of quarkonia states is expected in a de-confined medium due to the screening of long-range attractive potential. We are beginning studies of the ATLAS capabilities to identify ϒ states. The initial evaluation is that the stand-alone muon system will provide marginal resolution for a clear separation of the three states. However the use of the SCT and pixel detectors can enhance the mass resolution.

A detailed simulation of the ATLAS detector in the heavy ion environment is under way.  Figure 3 depicts one HIJING event for b=0 in the ATLAS detector. The detector geometry utilized is the same as used for proton-proton collision simulations.   The path for the simulation studies is very clear. For jet physics it is very important to establish what are the detection thresholds vis-à-vis the underlying soft background for one heavy ion event. It is also interesting to establish the detector's sensitivity to changes in the various jet parameters, e.g, cone radius. A detailed simulation for the di-muon invariant mass reconstruction is also under way.

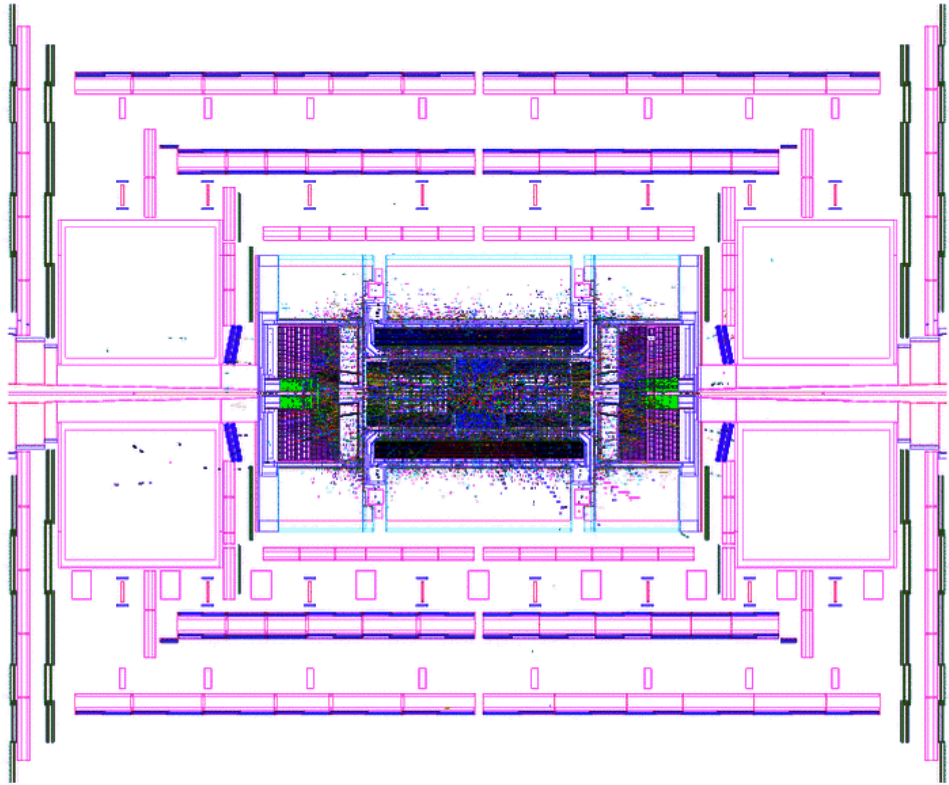

**Figure 3 - Simulated Pb+Pb event. The event generator used is HIJING and event simulated using GEANT-3.**

## 5. ATLAS Physics in Proton-Nucleus Collisions

The ATLAS detector provides an unprecedented opportunity to study proton-nucleus or light ion-nucleus collisions in a detector with both large acceptance and nearly complete coverage of the various final states that can result from perturbative QCD processes. Measurements of proton-nucleus collisions at the LHC will not only provide essential control of the hard processes that are expected to determine the initial conditions of heavy ion collisions, but these measurements can also address physics that is, in its own right, of fundamental interest. The full ATLAS detector will be functional for proton-nucleus collisions since they are not very different in occupancy (at most a factor of 5 larger than in proton-proton collisions for soft processes) than the expected 25 simultaneous proton-proton collisions per bunch crossing at LHC design luminosity

The ATLAS detector provides an existing facility in which critical, high-precision proton-nucleus measurements can be made with no additional hardware. The possible addition of a zero-degree calorimeter provides additional opportunities outlined below.

Much of the interest in proton-nucleus collisions at the LHC is focused on the perturbative production of gluons and the modifications of the gluon distribution in the nucleus at low x. Even at RHIC, the perturbative production of gluons is expected to play an important role in determining the initial conditions of a heavy ion collision [12,13]. At LHC energies, hard gluon production becomes even more important due to the growth of the gluon distribution at low x. For example the production of 20 GeV gluon jets through ordinary pQCD processes will involve initial-state gluons at an x of $\sim 2 \times 10^{-3}$. The more copiously produced lower $p_\perp$ gluon jets will probe even lower x values. At such low x, strong modifications of the nuclear gluon distribution are expected and understanding these modifications will be essential to understanding the initial conditions of heavy ion collisions at the LHC. At large $Q^2$, the modifications to the nuclear gluon distribution are best understood in the context of shadowing, and extensive pQCD calculations of the nuclear size and the x dependence of the shadowing modifications of the gluon distribution in heavy nuclei have been performed [14]. Studies of (e.g.) photon-jet and $b\bar{b}$ production in proton-nucleus collisions will allow these modifications to be measured providing both an essential test of the theoretical calculations and a important constraint on the initial conditions of heavy ion collisions. The large pseudo-rapidity coverage of ATLAS will allow the nuclear shadowing to be mapped out over an x range $\sim 10^{-2} < x < 10^{-4}$.

At lower $Q^2$, the evolution of the parton distributions in the nucleus becomes non-linear due to the large gluon density in the transverse plane [15]. The physics of saturation is controlled by a scale, $Q_s$, such that the growth in the gluon density in the nucleus with decreasing $k_\perp$ is cut off for $k_\perp \leq Q_s$. In the very small x range accessible at the LHC, $Q_s$ may be larger than 4 GeV [16], and since the gluon parton distribution $g(x,Q^2)$ is sensitive to gluons with $k_\perp^2 \leq Q^2$ [17] it should be possible to directly probe the saturated gluons by measuring hadron production over the $p_\perp$ range $2 \leq p_\perp \leq 4$ GeV/c.

The ATLAS inner detector has been designed to have good tracking efficiency in this momentum range over the pseudo-rapidity range $|\eta| \leq 2.5$. The ability to reconstruct jets in ATLAS over a larger pseudo-rapidity region and over the full azimuth would allow the potential contamination of the saturation physics from high $p_\perp$ jets to be reduced. One aspect of saturation that is not yet well understood is the affect on hard scattering processes for $Q^2 \sim Q_s^2$. It has been suggested that the gluon $k_\perp$ distribution could be significantly modified well above $Q_s$. If so, then ATLAS should be able to study the $Q^2$ evolution of the saturation effects using $\gamma$-jet, $b\bar{b}$, and jet-jet measurements.

Another important physics topic that can be addressed by proton-nucleus measurements in ATLAS is the applicability of factorization in hard processes involving nuclei. No detector that has studied either nuclear deep inelastic scattering or proton-nucleus collisions has been able to simultaneously study many different hard processes to explicitly demonstrate that factorization applies. In ATLAS, we will have the opportunity to study single jet and jet-jet events, photon-jet events, heavy quark production, Z and $W^\pm$ production, Drell-Yan production of di-leptons, etc. and probe in detail the physics of jet fragmentation. With the wide variety of available hard processes that all should be determined by the same parton distributions, we should be able to demonstrate clearly the success of factorization for $Q^2 >> Q_s^2$.

Detailed studies of the behavior of jet fragmentation as a function of pseudo-rapidity could be used to determine if the jet fragmentation is modified by the presence of the nucleus or its large number of low-x gluons. The observation or lack thereof of modifications to jet fragmentation could provide sensitive tests of our understanding of formation time and coherence in the re-dressing of the hard scattered parton and the longitudinal spatial spread of low-x gluons in a highly Lorentz contracted nucleus.

Yet another interesting problem that can be well studied in proton-nucleus collisions at the LHC is that of double hard scattering events. These are events in which, for example, two separate partons in the proton undergo a hard scattering in the nucleus. Such events have been observed in p-pbar collisions at the Tevatron [18]. In addition to providing a sensitive test of QCD, they directly probe parton correlations in the proton [19]. It has been argued that when these events take place in a nuclear target, the correlations from the partons in the nucleus are relaxed, so by comparing p-p and p-A data on double-hard scattering, the understanding of parton correlations in the nucleon may be dramatically improved [20].

## 6. ATLAS Physics in Ultra-Peripheral Nuclear Collisions

At the LHC, the highly Lorentz contracted electric field of the ions can be viewed as a source of high energy photon-photon and photon-nucleon collisions, that are physics competitive with electron-positron colliders and electron deep inelastic scattering at HERA [21]. The ATLAS program with Ultra-Peripheral collisions will extend the systematic study of hadron structure at HERA to higher energies and to nuclear targets.

Recently, there has been a lot of activity in the field of ultra-peripheral collisions due, in part, to the realization that at the LHC γ-nucleus collisions will occur at up to several hundred TeV equivalent laboratory energy (in the nucleus rest frame). After the Erice meeting [21] a case for this program was formulated in a white paper [22]. Subsequently, a working group was formed to write a CERN yellow report dealing with the theoretical issues and the potential for relevant measurements with the LHC experiments (ATLAS, ALICE and CMS).

Approximately half a dozen topics are on the "short list" of the γ–γ program as a result of the Erice statement but it is likely that, at the LHC, an important goal will be the study of partonic distributions as probed by heavy quark production by high energy photons. Also hard diffraction of photo-produced vector mesons is of interest. Of course, a feature of our experiment is that the photon $q^2$ range is limited to $\sim 1/R_{nucleus}^2$ in order to take advantage of the large ($Z^2$) enhancement in the cross section. Quasi-real photons are emitted coherently from the nucleus. A more detailed discussion of the equivalent photon spectra with comparisons to the electron-ion collider and RHIC can be found elsewhere [23].

Heavy quark photo-production at LHC energies was studied by Gellis and Peshier [24] via both photon-gluon fusion and diffractive scattering of the hadronic component of the photon. Their calculation was carried out within the framework of the Colored Glass Condensate). Another approach to the same calculation [25] uses a different description of the parton distributions. The measurements by ATLAS of heavy quark photoproduction will certainly be pertinent in distinguishing between these and other calculations which can be found in the literature.

The ATLAS detector fulfills the prerequisites for a program of physics in this field in that it has hermetic coverage with tracking and high quality calorimetry over full azimuth and out to pseudo-rapidity of $\eta<4.9$ units. Whereas in e+e- and e-hadron colliders, the photon energy is tagged via the recoil electron, in ultra-peripheral heavy ion collisions gamma and Pomeron mediated interactions are identified via rapidity gaps (regions of phase space free of particles as expected for colorless exchange) and kinematic quantities. The tools that are currently available, including those which we have developed in analyzing electromagnetic interactions at RHIC, are adequate for eliminating the obvious hadronic backgrounds.

It is sometimes said that the photon energy cannot be determined without some sort of "tag" by the particle that radiates the photon. For this reason an equivalent program using p-p interactions (where the leading proton can at least marginally, in principle be detected) has been discussed[26]. On the other hand, at large enough center of mass energy energies, the $E_T$ weighted rapidity distribution of particles in such an event can provide equivalent or better energy of the incident photon as demonstrated by the hadronic diffraction publications by CDF and D0. Electromagnetic and Pomeron mediated interactions are very similar kinematically and the comparison may be fruitful.

## 7. Conclusions

ATLAS is a detector with enormous physics capabilities. The relativistic heavy ion collider (RHIC) at Brookhaven National Laboratory has produced a great deal of excitement in its two first years of running. The ATLAS nuclear physics program represents a measurement of the heavy ion physics excitation function up to un-precedented energies. We are working towards a full proposal with complete ATLAS physics simulations over the next year.